\documentclass{article}
\usepackage{spconf,amsmath,graphicx,hyperref,enumitem}
\usepackage{spconf,amsmath,graphicx}
\usepackage{todonotes}
\usepackage{soul}
\usepackage{hyperref}
\usepackage{xspace}
\usepackage{blindtext}

\definecolor{TodoColor}{rgb}{1,0.7,0.6}

\usepackage{xstring}
\usepackage{seqsplit}
\usepackage{placeins}

\newcommand{\capitalfirst}[1]{%
    \StrLeft{#1}{1}[\firstletter]%
    \StrGobbleLeft{#1}{1}[\restofword]%
    \MakeUppercase{\firstletter}\restofword%
}

\newcommand{\capitalizehyphenated}[1]{%
    \StrCut{#1}{-}{\firstpart}{\restpart}%
    \capitalfirst{\firstpart}%
    \IfStrEq{\restpart}{}{}{ \capitalizehyphenated{\restpart}}
}

\usepackage[capitalise]{cleveref}
\crefname{figure}{Figure}{Figures}
\crefname{table}{Table}{Tables}
\crefname{appendix}{Appendix}{Appendices}

\usepackage{multirow}

\usepackage[table]{xcolor}
\definecolor{taskgreen}{RGB}{180, 220, 180}
\definecolor{taskblue}{RGB}{180, 210, 230}
\definecolor{taskorange}{RGB}{255, 210, 170}
\definecolor{taskpink}{RGB}{255, 185, 200}

\renewcommand{\paragraph}[1]{\medskip\noindent\textbf{#1}\quad}

\usepackage{booktabs}
\usepackage{tabularx}
\usepackage{arydshln}

\makeatletter
\def\adl@drawiv#1#2#3{%
  \hskip.5\tabcolsep
  \xleaders#3{#2.5\@tempdimb #1{1}#2.5\@tempdimb}%
    #2\z@ plus1fil minus1fil\relax
  \hskip.5\tabcolsep
}
\newcommand{\cdashlinelr}[1]{%
  \noalign{\vskip\aboverulesep
           \global\let\@dashdrawstore\adl@draw
           \global\let\adl@draw\adl@drawiv}%
  \cdashline{#1}%
  \noalign{\global\let\adl@draw\@dashdrawstore
           \vskip\belowrulesep}%
}
\makeatother

\usepackage{amssymb}
\usepackage{orcidlink}
\usepackage{etoolbox}
\hypersetup{
  urlcolor     = blue, 
  linkcolor    = blue, 
}

\usepackage{xspace}
\newcommand{\VA}{\textsc{Voice Arena Goal Dataset}\xspace}

\setlist{noitemsep,left=0mm,topsep=1mm}
\usepackage{xfrac}
\usepackage{bbm}
\usepackage{bm}

\makeatletter 
\patchcmd{\thebibliography}
  {\advance\leftmargin\labelsep}
  {\setlength{\itemsep}{1pt plus 0.2pt}
   \setlength{\parsep}{1pt}
   \setlength{\parskip}{1pt}
   \advance\leftmargin\labelsep}
  {}
  {}
\title{Calibrating LLM Judges for Human and AI Conversations}

\name{\parbox{\dimexpr\textwidth-2\tabcolsep\relax}{\centering
Maike Z\"{u}fle\,\orcidlink{0009-0001-7238-7705}\textsuperscript{*,1},
Patr\'{i}cia Schmidtov\'{a}\,\orcidlink{0009-0008-5516-798X}\textsuperscript{*,2},
Vil\'{e}m Zouhar\,\orcidlink{0000-0001-9874-2069}\textsuperscript{3},
Shree Harsha Bokkahalli Satish\,\orcidlink{0009-0000-0554-7265}\textsuperscript{4},\\
Erica Cooper\,\orcidlink{0000-0002-2978-2793}\textsuperscript{5},
Nicholas Sanders\,\orcidlink{0009-0007-0925-1606}\textsuperscript{7},
Shobhit Banga\,\orcidlink{0009-0001-0896-5902}\textsuperscript{6},
Vaibhav Nalawade\,\orcidlink{0009-0006-1871-1128}\textsuperscript{6},
Manmeet Kaur\,\orcidlink{0009-0005-7144-0163}\textsuperscript{6},\\
Jan Niehues\,\orcidlink{0000-0002-4231-6543}\textsuperscript{1},
Markus M\"{u}ller\,\orcidlink{0009-0000-1701-0019}\textsuperscript{1},
Ond\v{r}ej Klejch\,\orcidlink{0000-0001-5495-967X}\textsuperscript{7}
}\thanks{\textsuperscript{*}Equal Contribution}}

\address{\parbox{\dimexpr\textwidth-2\tabcolsep\relax}{\centering
\textsuperscript{1}Karlsruhe Institute of Technology, Germany
\textsuperscript{2}Charles University, Czech Republic
\textsuperscript{3}ETH Zurich, Switzerland
\textsuperscript{4}KTH Royal Institute of Technology, Sweden
\textsuperscript{5}NICT, Japan
\textsuperscript{6}Voice Arena, USA
\textsuperscript{7}University of Edinburgh, UK\\
\href{mailto:maike.zuefle@kit.edu}{\texttt{maike.zuefle@kit.edu}}
}}
\begin{document}
%
\maketitle
\begin{abstract}
Measuring how successful a conversation is remains difficult, even for humans judging spoken dialogue. We evaluate state-of-the-art LLMs as pointwise and pairwise judges of conversational success on CANDOR, finding pointwise scoring correlates moderately with human ratings, while pairwise comparison suffers from long transcripts and positional bias. Since this leaves judge scores incomparable across models, we propose a small anchor set and a calibration function that calibrates any judge onto a shared, interpretable scale. We further release the \VA (VA), 200 task-oriented human-AI and human-agent conversations with pairwise annotations, revealing a substantial gap between current judges and human-level discrimination. Using VA, we test whether CANDOR-fitted calibration transfers to human-AI conversations, finding it brings judges onto a shared scale despite never observing VA during fitting.
\end{abstract}

\section{Introduction}
\label{sec:intro}

Conversational speech models are becoming increasingly popular~\cite{hu2025salmduplexefficientdirectduplex, défossez2024moshispeechtextfoundationmodel, roy2026personaplexvoicerolecontrol,zufle-etal-2026-f}, and technical progress is fast. However, it remains difficult to evaluate how good a conversation is, even though we wish to quantify it to develop better conversational systems. How \textit{successful} is a conversation? Even for conversations between humans only, this is difficult to judge.

Recent work introduced the CANDOR dataset~\cite{reece2023candor} where humans judged how successful their own conversations were. In practice, collecting such self-reports is not feasible for every dialogue that a model or human produces. LLM judges offer a more scalable alternative, and prior work has shown they can align well with human judgments on related tasks~\cite{zheng2023judging,liu2023g}.

In this paper, we evaluate state-of-the-art models as judges of conversational success. We find that direct, pointwise scoring achieves moderate correlation with human ground truth, but scores are not comparable across judges: different models produce scores on different scales, and since new models are introduced constantly, different works rarely use the same judge, making cross-paper comparison unreliable~\cite{retkowski-etal-2025-summarizing}. 

One natural fix is pairwise comparison, where a judge is shown two conversations and asked which is more successful, avoiding the need for a shared absolute scale. However, we find pairwise judging is not a reliable solution in practice: conversations can run 25 minutes or longer, and presenting two full conversations to a judge at once exceeds the context window of many models, some of which do not support inputs of this length at all. Even when input is reduced to transcript-only text to fit within context limits, positional bias, where a judge's preference depends on which conversation is presented first, further undermines pairwise reliability \cite{zheng2023judging, wang2024large,bokkahalli2025voice}.

We therefore propose using a subset of the CANDOR dataset as an \textit{anchor dataset}, against which any judge can be calibrated, together with a transformation function that maps a judge's raw scores onto a shared, interpretable scale.

To extend judging to human-AI conversations, the setting voice agents are built for, we introduce the \VA (VA), 200 conversations in which a human discusses a topic with either a human agent or one of four state-of-the-art models. We collect human-annotated rankings across these conversations and release the dataset for public use.
We apply our CANDOR-calibrated judges to VA and assess whether the calibration transfers, bringing these judges' scores onto a shared, comparable scale. 

The contributions of this paper are as follows\footnote{Code: \href{https://github.com/MaikeZuefle/conv-judge-calib}{conv-judge-calib}, data: \href{https://huggingface.co/datasets/VoiceArena/Goal-Dataset_en_in}{\VA}}:
\begin{itemize}
    \item We evaluate state-of-the-art models judging human-human dialogues, comparing pointwise and pairwise judging.
    \item We propose an anchor dataset and a transformation function to calibrate LLM judges onto a shared, comparable scale.
    \item We release the \VA, a human-AI dataset with human-annotated pairwise comparisons of conversations, showing that current LLM judges struggle to discriminate human-AI conversational success.
\end{itemize}

\section{Background and Related work}\label{sec:bck-rw}

\paragraph{Evaluation of non-task-oriented conversations.}
A conversation admits many appropriate responses, and their quality depends on many properties that accumulate over turns. Lexical-overlap metrics may therefore correlate poorly with human judgments~\cite{liu2016not}, motivating reference-free and dialogue-level metrics such as USR, FED, and FineD-eval~\cite{mehri2020usr,mehri2020unsupervised,zhang2022fined}.
\begin{table*}[t]
\centering
\small
\begin{tabular}{lllcccccc}
\toprule
\multirow{2}{*}{\textbf{Model}} & \multirow{2}{*}{\textbf{Mod.}} & \multirow{2}{*}{\textbf{Prompt}} & \multirow{2}{*}{$\bm{\rho}$} & \multirow{2}{*}{$\bm{\alpha}$} & \multicolumn{2}{c}{\textbf{PA (Gen.)}} & \multicolumn{2}{c}{\textbf{PA (HSC$\times$LSC)}} \\
\cmidrule(lr){6-7} \cmidrule(lr){8-9}
 & & & & & \textbf{Pointwise} & \textbf{Pairwise} & \textbf{Pointwise} & \textbf{Pairwise} \\
\midrule
Length baseline                  & text     & --                 & --    & --    & 61.3\% & 61.3\% & 83.2\% & 83.2\% \\
\midrule
\multirow{2}{*}{Qwen2.5-Omni-7B}  & text   & successful conversation?            & 0.24  & 0.07  & \textbf{66.4\%} & 52.2\% & \textbf{94.3\%} & 55.0\%\\
                                  & audio  & CoT: CANDOR questions      & 0.09  & 0.07  & 52.1\% & 52.5\%  & 67.0\% & 62.1\% \\
\multirow{2}{*}{Phi-4-Multimodal} & text   & CoT: summary of conv.       & 0.21  & 0.02  & 63.5\% & 50.0\% & 73.4\% & 50.0\%\\
                                  & audio  & CoT: summary of conv.       & \textbf{0.32}  & 0.24  & 66.3\% & 50.0\% & 93.6\% & 50.0\% \\
\multirow{2}{*}{Qwen3-Omni-30B}   & text   & enjoyable conversation?                & 0.24   & 0.17   & 59.8\%    & 61.3\% & 71.4\%    & 80.0\% \\
                                  & audio  & CoT: summary of conv.               & 0.20   & 0.10   & 62.1\%    & 59.3\% & 86.6\%    & 76.6\%  \\
Qwen3.5-27B                       & text   & successful / enjoyable conv.?            & 0.28  & \textbf{0.25}  & 62.8\% & \textbf{64.8\%}  & 90.3\% & \textbf{91.0\%} \\
\bottomrule
\end{tabular}
\caption{CANDOR results. Spearman's $\rho$ ($\uparrow$) and Krippendorff's $\alpha$ ($\uparrow$) are computed against the PCS score. For pointwise judges, pairwise accuracy (PA) is derived from the individual scores. We report the best prompt for each model and modality. 
}
\label{tab:candor-results}
\end{table*}

Human evaluation protocols also offer a complementary lens. ACUTE-EVAL compares speakers across complete dialogues~\cite{li2019acute}, and~\cite{smith2022human} show that pairwise dialogue judgments can expose differences that emerge over several turns, although no protocol is uniformly resilient. 

We build on this observation by comparing pointwise and pairwise LLM judges, using participant self-reported conversational success rather than chatbot quality as reference.

\paragraph{LLMs as evaluators/judges.}
LLM judges can align well with human preferences on open-ended generation tasks~\cite{zheng2023judging,liu2023g}, but their validity varies across tasks and evaluated properties \cite{wang2024large,bokkahalli2026voice,bavaresco2025llms}. They also exhibit position, verbosity, and self-preference biases~\cite{zheng2023judging,wang2024large,bokkahalli2025voice}. Speech adds information that is unavailable in transcripts, including prosody, timing, affect, and speaker characteristics. However, it also introduces new failure modes. SpeechLLM judges can approach human agreement for speaking-style evaluation~\cite{chiang2025audio}, and AudioJudge reports strong correlation with human speech preferences when lexical, acoustic-quality, and paralinguistic assessments are decomposed and ensembled~\cite{manakul2026audiojudge}.

\section{Judging Conversational Success}\label{sec:exp}
We evaluate state-of-the-art LLMs as judges of conversational success under two elicitation strategies: \textit{pointwise}, where a judge assigns a single conversation an absolute score, and \textit{pairwise}, where a judge is shown two conversations and asked which is more successful. All judges are used zero-shot, without fine-tuning, to avoid overfitting them to a single corpus.

\paragraph{Pointwise Judges.}
We prompt four models to assign each conversation a single success score from 0--10: three audio-capable models given either the raw audio or its transcript (Qwen2.5-Omni-7B~\cite{xu2025qwen25omnitechnicalreport}, Phi-4-Multimodal-Instruct~\cite{microsoft2025phi4minitechnicalreportcompact}, and Qwen3-Omni-30B~\cite{xu2025qwen3omnitechnicalreport}),\footnote{We also conducted preliminary experiments with Audio-Flamingo~\cite{ghosh2026audioflamingonextnextgeneration},  but were unable to reliably extract scores from its responses under our setup.}
and a text-only model judging transcripts alone (Qwen3.5-27B~\cite{qwen35}). Beyond a direct rating prompt, we test several chain-of-thought (CoT) variants that ask the model to first summarize the conversation, answer a fixed set of CANDOR survey-derived questions, or produce question-relevant partial summaries, before emitting a final score, following the intuition that intermediate reasoning steps can improve LLM-judge reliability~\cite{zheng2023judging}. 

\paragraph{Pairwise Judges.}
Following pairwise preference judgment \cite{park2024paireval},  judges are shown two conversations and asked to select the more successful one, instead of scoring each separately.
We use the same models and prompting strategies as in the pointwise experiment. To obtain pairwise judgments of audio, we chunk the conversations into sliding windows, ranking the snippets. The conversation with the most winning snippets wins the comparison.
We also report a content-blind baseline by always choosing the longer transcript.

\subsection{Experimental Setup}
\paragraph{Data.}
CANDOR~\cite{reece2023candor} comprises 1656 recorded conversations between strangers instructed to talk for at least 25 min, with extensive post-conversation surveys and transcripts. We use survey items such as enjoyment, conversational success, perceived affect, and quality of conversations as ground-truth labels for judging. Following~\cite{withanage2026acoustic}, we derive a continuous 0--1 \textit{Perceived Conversation Success} (PCS) score from these survey responses for each conversation. Of the 1656 recordings, 20 lack a computable PCS score and are excluded, leaving 1636 conversations for evaluation. 
Following the grouping of~\cite{reece2023candor}, we additionally bin conversations into three success categories based on their PCS score: High-Success Conversations (HSC, $n=91$), Medium-Success Conversations (MSC, $n=1510$), and Low-Success Conversations (LSC, $n=35$). 

For the pairwise evaluation, we construct an exhaustive HSC$\times$LSC test set (91$\times$35=3185 pairs), presenting each pair in both orderings to separate genuine preference from positional bias. As a more difficult alternative, we construct a set of conversation pairs where the absolute difference between any given pair is $>$0.1 PCS. We discard 6 long outliers and then sample 1000 conversation pairs (\textit{Gen.}).

\paragraph{Evaluation.}
For pointwise judges, we compute Spearman's $\rho$ and Krippendorff's $\alpha$ between the extracted score and the continuous PCS ground truth across all 1636 conversations. To compare pointwise and pairwise judges, we additionally derive a \textit{pairwise accuracy} for pointwise judges.

\subsection{Results of the Judges}

\Cref{tab:candor-results} summarizes the judge performance on CANDOR. Phi-4-Multi\-modal (audio, $\rho=0.32$) and Qwen3.5-27B (text, $\rho=0.28$) show the strongest pointwise correlations with PCS.
On the easier HSC$\times$LSC pairwise test, pointwise-derived accuracy reaches 93.6\% and 90.3\%, matching or exceeding the length baseline (83.2\%) and the explicit pairwise judge Qwen3.5-27B (91.0\%), suggesting pairwise elicitation offers no clear advantage.
On the general comparison test, models score below 70\%, close to the 61.3\% length baseline, showing that separating clear cases is easy but discriminating within the broad middle remains hard regardless of elicitation strategy. Phi-4 and Qwen2.5 show a strong positional bias in the pairwise setting, picking the first-presented conversation around 90\% of the time. We find no consistent modality advantage across models.

\section{Calibrating Future Judges}

\begin{figure}[htbp]
    \centering
    \includegraphics[width=\linewidth]{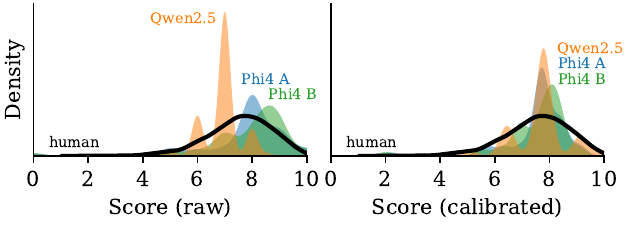}
    
    \vspace{-3mm}
    \caption{Raw (left) and calibrated (right) distributions of scores based on LLM judges and human annotators.}
    \label{fig:score_distribution}
\end{figure}

We naturally desire that conversation scores are objective and comparable, but a PCS score computed with a new LLM judge model is not soundly comparable to human scores or scores by previous LLM judges. To alleviate this, we describe an efficient calibration procedure which requires only little computational overhead. As a result, the score distributions from all LLM judges match those of human annotators on CANDOR which aids comparability of scores. Consider \Cref{fig:score_distribution} (left), where different LLM judges produce different score distributions (e.g., Qwen2.5 consistently underestimates the true score). Upon our proposed calibration, described in this section, these distributions are more aligned.

\paragraph{Problem formalization.}
Each conversation $x$ has a conversational success score either by human $c_h(x)$ or some LLM judge model $m$: $c_m(x)$.
We wish to find such a calibration function $t$ that minimizes the Wasserstein distance (optimal transport distance) between $c_h$ and $t \circ c_m$ across a natural distribution of conversations $\mathbb{X}$.
The search for $t$ needs to be practical, so we want to select an anchor set $A \subseteq \mathrm{support}(\mathbb{X})$ where we have access to true human scores $c_h(x_a)$ and LLM judge predictions $c_m(x_a)$ for $x_a \in A$.

\paragraph{Selecting a calibration function.}
We optimize a transformation $t$ on the anchor set.
To prevent overfitting and remain practical, we consider only affine transformations $t(c)=\theta_0 + \theta_1 \cdot c$.
We wish to improve the comparability of judge scores but not at the cost of lowering correlations with human judgments.
With $\theta_1 > 0$, the function is strictly increasing, such that ranking correlations are preserved, $\rho(c_h, c_m) = \rho(c_h, t\circ c_m)$. The functions we consider are listed in \cref{tab:calibration}. We fit the parameters by either minimizing the mean-squared error (MSE) or the Wasserstein distance. Min/max- and $\mu$/$\sigma$-matching align the judge's range or mean and standard deviation (SD) with the human anchor scores.

The \textbf{anchor} needs to be representative of $c_h(\mathbb{X})$.
We consider randomly selecting an anchor $A_\mathrm{random}$ of a given size, and an optimized anchor $A_\mathrm{optimized}$, chosen via random search over 1000 candidate subsets to minimize the mean Wasserstein loss on the other models. We exclude the model it is later evaluated on, so the anchor is not tailored to that judge's own score distribution.
We compare several transformations under both anchors at anchor sizes 8, 32, and 128, evaluating each on conversations held out from the anchor.

\begin{table}[t]
\centering \small
\setlength{\tabcolsep}{2.5pt}
\begin{tabular}{l ccc @{\hspace{15pt}} ccc}
\toprule
\multirow{2}{*}{\bf Method}
& \multicolumn{3}{c@{\hspace{10pt}}}{$\mathbf{A}_\textbf{random}$}
& \multicolumn{3}{c}{$\mathbf{A}_\textbf{optimized}$} \\[-0.4em]

& \bf\tiny 8 & \bf\tiny 32 & \bf\tiny 128
& \bf\tiny 8 & \bf\tiny 32 & \bf\tiny 128 \\[-0.4em]
\midrule
Identity
& \cellcolor{black!0}0.70 & \cellcolor{black!0}0.70 & \cellcolor{black!0}0.71
& \cellcolor{black!0}0.70 & \cellcolor{black!1}0.69 & \cellcolor{black!1}0.69 \\
MSE Constant
& \cellcolor{black!0}0.79 & \cellcolor{black!5}0.66 & \cellcolor{black!10}0.62
& \cellcolor{black!14}0.60 & \cellcolor{black!16}0.58 & \cellcolor{black!13}0.61 \\
MSE Linear
& \cellcolor{black!0}0.81 & \cellcolor{black!10}0.63 & \cellcolor{black!11}0.62
& \cellcolor{black!11}0.62 & \cellcolor{black!15}0.59 & \cellcolor{black!15}0.59 \\
MSE Affine
& \cellcolor{black!0}0.73 & \cellcolor{black!0}0.77 & \cellcolor{black!0}0.73
& \cellcolor{black!8}0.64 & \cellcolor{black!16}0.58 & \cellcolor{black!8}0.64 \\
Wass. Affine
& \cellcolor{black!0}0.70 & \cellcolor{black!30}0.48 & \cellcolor{black!38}0.42
& \cellcolor{black!22}0.54 & \cellcolor{black!40}0.41 & \cellcolor{black!38}0.42 \\
$\sfrac{\mathrm{min}}{\mathrm{max}}$-matching
& \cellcolor{black!0}0.95 & \cellcolor{black!0}1.11 & \cellcolor{black!0}0.96
& \cellcolor{black!0}0.80 & \cellcolor{black!0}0.81 & \cellcolor{black!0}0.86 \\
$\sfrac{\mu}{\sigma}$-matching
& \cellcolor{black!9}0.64 & \cellcolor{black!25}0.51 & \cellcolor{black!34}0.45
& \cellcolor{black!28}0.49 & \cellcolor{black!30}0.48 & \cellcolor{black!32}0.46 \\
\bottomrule
\end{tabular}
\caption{Average Wasserstein distance ($\downarrow$) between human and transformed LLM judge distributions for random vs. optimized anchors of varying size, evaluated on held-out conversations outside the anchor, across the judges from \cref{tab:candor-results}.}
\label{tab:calibration}
\end{table}

\paragraph{Calibration results and recommendations.}
Results are shown in \Cref{tab:calibration}.
Wasserstein Affine and $\mu$/$\sigma$-matching are consistently the best-performing methods. Wasserstein Affine is ahead at sizes 32 and 128, while $\mu$/$\sigma$-matching performs better only at anchor size 8. 
Optimized anchor selection improves over random selection, most noticeably at small anchor sizes, and the two nearly converge by size 128.

We therefore select 32 as our anchor size, a middle ground between the noise of small anchors and the labeling cost of  large ones. We provide a fixed, optimized 32-item anchor set, optimized over all available judges. This anchor set averages 33.3 min per conversation (SD 10.9, range 26.4--90.1), close to the full CANDOR corpus's mean of 34.1 min, and has a mean PCS score of 0.785 (SD 0.094, range 0.641--0.996).
When introducing a new judge, we recommend running it on this anchor and calibrating it with closed-form $\mu$/$\sigma$-matching.

\section{Transfer to Human-AI Conversations}

\subsection{Introducing the \VA}
To evaluate conversational judges in human-AI settings, we introduce the \VA (VA), a dataset of 200 task-oriented voice conversations (17.75 hours total). Each conversation places a human caller against one of five conditions: four state-of-the-art conversational AI systems (Gemini~\cite{google_gemini_live_api}, Grok~\cite{xai2025grokvoice}, OpenAI~\cite{openai2024realtime}, and Inworld~\cite{inworld_realtime_api}) and a human agent baseline. Conversations are drawn from two task scenarios, \textit{booking a flight for two people} and \textit{rescheduling one passenger}, with 100 conversations per scenario. Within each scenario, conversations are further organized into 20 task instances (e.g., a specific route and date), each instance recorded once under all five conditions, yielding a balanced 5-way comparison per instance. 
For each task instance, 1--6 (average 4) human raters compared the five conversations pairwise, judging both \textit{task capability} (did the agent successfully complete the task) and \textit{humanness} (how human-like the agent sounded). This yields 1606 pairwise judgments across 397 unique conversation pairs. We use the majority vote as the gold ranking and for each criterion only take the pairs that are not tied (n$=$260 and 299, respectively).

\subsection{Discrimination Accuracy}
\begin{table}[t]
\centering
\small
\setlength{\tabcolsep}{5.2pt}
\begin{tabular}{llcc}
\toprule
\textbf{Model} & \textbf{Mod.} & \textbf{PA Task}\ (\%) & \textbf{PA Human} (\%) \\
\midrule
Duration baseline    & --    & $65.0\phantom{{\scriptstyle\pm 0.0}}$ & $60.4\phantom{{\scriptstyle\pm 0.0}}$ \\
Human ceiling        & --    & $84.2\phantom{{\scriptstyle\pm 0.0}}$ & $86.7\phantom{{\scriptstyle\pm 0.0}}$ \\
\midrule
\multirow{2}{*}{Phi-4-Multimodal}     & audio & $49.5\phantom{{\scriptstyle\pm 5.0}}$ & \phantom{$^\dagger$}$67.2^\dagger\phantom{{\scriptstyle\pm 2.2}}$ \\
                              & text  & $54.4{\scriptstyle\pm 5.0}$ & $52.9{\scriptstyle\pm 2.2}$ \\
\multirow{2}{*}{Qwen2.5-Omni-7B} & audio & $46.9\phantom{{\scriptstyle\pm 8.6}}$ & $46.5\phantom{{\scriptstyle\pm 7.6}}$ \\
                              & text  & \phantom{$^\dagger$}$64.7^\dagger{\scriptstyle\pm 8.6}$ & $60.0{\scriptstyle\pm 7.6}$ \\
Qwen3.5-27B                   & text  & $52.2{\scriptstyle\pm 5.0}$ & $54.2{\scriptstyle\pm 1.2}$ \\
\multirow{2}{*}{Qwen3-Omni-30B} & audio & $38.4\phantom{{\scriptstyle\pm 1.6}}$ & $37.9\phantom{{\scriptstyle\pm 0.3}}$ \\
                                  & text  & $43.1{\scriptstyle\pm 1.6}$ & $47.5{\scriptstyle\pm 0.3}$ \\
\midrule
Phi4$\to$Qwen3.5     & audio & $47.2\phantom{{\scriptstyle\pm 0.0}}$ & \phantom{$^\dagger$}$59.3^\dagger\phantom{{\scriptstyle\pm 0.0}}$ \\
Qwen2.5$\to$Qwen3.5   & audio & $52.8\phantom{{\scriptstyle\pm 0.0}}$ & $51.5\phantom{{\scriptstyle\pm 0.0}}$ \\
\bottomrule
\end{tabular}
\caption{\VA pairwise accuracy ($\uparrow$), best prompt. Text values are reported as mean $\pm$ half the range between Parakeet~\cite{sekoyan2025canary1bv2parakeettdt06bv3efficient} and Whisper Large V3~\cite{radford2022robustspeechrecognitionlargescale} ASR transcription. $^\dagger$ marks accuracy significantly above chance (95\% CI excludes 50\%). Model1$\to$Model2 rows use the named model to describe the audio, then judge with text. Human ceiling: leave-one-rater-out agreement.
}\vspace{0.3cm}
\label{tab:voicearena-results}
\end{table}

 \begin{table}[t]
\centering \small
\begin{tabular}{l ccc}
\toprule
\textbf{Dimension} & \hspace{4mm}\textbf{Raw}\hspace{4mm} & \textbf{$\bm{\mu}$/$\bm{\sigma}$-match.} & \textbf{Wass. Affine} \\
\midrule
Overall (success)  & 1.44 & 0.29 & 0.19 \\
Humanness          & 3.08 & 0.61 & 0.30 \\
Task capability    & 2.31 & 0.84 & 0.30 \\
\bottomrule
\end{tabular}
\caption{Mean pairwise Wasserstein distance ($\downarrow$) between judges' \VA score
distributions, before and after calibration fit on CANDOR's anchor set. Both methods bring judges onto a shared scale.}
\label{tab:calibration-transfer}
\end{table}

\cref{tab:voicearena-results} reports pairwise accuracy (PA) on VA. A simple duration baseline (shorter is better) already reaches $65.0\%$ (task) and $60.4\%$ (humanness), while human agreement is $84.2\%$ and $86.7\%$. No LLM judge closes this gap: most score at or below the duration baseline. Only three variants are significant. This shows current LLM judges are not yet reliable discriminators of human-AI conversational success, motivating VA as a benchmark for this open problem.

\subsection{Calibration-Transfer Test}
The calibration function $t$ is fit exclusively on CANDOR and never observes VA.  Since CANDOR provides only an overall success score, the same CANDOR-fit calibration is applied to all three VA dimensions. We test whether $t$ achieves the property it is designed for, making different judges' scores comparable, on a domain it never saw. We measure the average pairwise Wasserstein distance between judges' VA score distributions, before and after calibration, for the overall success score and for the humanness and task-capability dimensions separately (\Cref{tab:calibration-transfer}). Raw distances are large (1.44--3.08), reflecting judges' different scales. Calibration collapses this by 3--10$\times$ under both methods, showing the CANDOR-fit calibration transfers to bring judges onto a shared scale on VA.

\section{Discussion and Conclusion}\label{sec:concl}
Judging conversations, especially between humans and AI, is hard, and different LLM judges are not calibrated to the same scale, making their scores incomparable. We advance this field by introducing the \VA, a new benchmark for testing conversational judges in human-AI settings, and a calibration method that brings judges from different models onto a shared, comparable scale.

\clearpage
\section{Acknowledgments}
This work was supported by JSALT 2026 at JHU with funds from NSF CCRI Grant No. 2120435, Google DeepMind, JHU HLTCOE, JHU AI2AI and ACL, and received funding from the European Union’s Horizon research and innovation programme under grant agreement No 101135798, project Meetween (My Personal AI Mediator for Virtual MEETtings BetWEEN People). 
PS was supported by CUNI projects GAUK 252986 and SVV 260 698 and co-funded by the European Union (ERC, NG-NLG, 101039303).

Generative AI tools were used to assist with editing and grammar checking of the manuscript, as well as for coding and plotting. All scientific content, analyses, and conclusions were developed and verified by the authors.

\bibliographystyle{IEEEtran}
\bibliography{strings,refs}

\end{document}